# Does "Fans Economy" Work for Chinese Pop Music Industry ?

Hao Wang
Beijing, China
haow85@live.com

*Abstract—* China has become one of the largest entertainment markets in the world in recent years. Due to the success of Xiaomi , many Chinese pop music industry entrepreneurs believe "Fans Economy" works in the pop music industry. "Fans Economy" is based on the assumption that pop music consumer market could be segmented based on artists. Each music artist has its own exclusive loyal fans. In this paper, we provide an insightful study of the pop music artists and fans social network . Particularly, we segment the pop music consumer market and pop music artists respectively. Our results show that due to the Matthew Effect and limited diversity of consumer market, "Fans Economy" does not work for the Chinese pop music industry.

*Keywords—Fans Economy; Popular Music; Social Network Analysis; Marketing; Market Segmentation; Community Detection*

## I. INTRODUCTION

With the continuous rapid growth of the Chinese economy, China has become one of the largest entertainment market in the world. [show some statistics] The term "Fans Economy" has been popular (especially in 2015) in China due to the success of the Chinese mobile phone brand Xiaomi. Xiaomi introduces high-end smart phones with low prices to create loyal fans of their products. Some popular music companies and websites in China also wants to create fans of the artists they newly produce. Meantime, many firms wish to sell products related to pop music artists to earn easy money. Following the latest trend in China , many of them have become firm believers of the "Fans Economy" assumption that for a specific music artist, there exists a group of fans that are loyal to the artist only. Some entertainment companies want to "sell" their music artists like Xiaomi sells its smart phones.

From marketing theory, we know that market segmentation is an inseparable part of marketing process. Unless the firm is in a perfect competitive market serving the mass population, like agriculture, the firm needs to know to which people they want to sell their products. If "Fans Economy" really works in the Chinese popular music industry, the consumer market must first be segmentable based on artists. In other words, if an entertainment company considers its newly introduced artist as a product and if there really could be a fans group loyal exclusively to that artist, then the fans should be separable into different groups according to different music artists. Similarly, if firms want to sell products related to a specific artist based on the assumption there exists loyal fans for that artist only, the buyers of such products should be different from buyers of another artist. On the other hand, if the music market is segmentable, music artists should also be segmentable based on their particular music styles based on fans.

In this paper we would like to debunk the theory that "Fans Economy" work for the Chinese popular music industry. We point out that with pop music artists taken as products , the consumer market exhibit serious Matthew Effect and very limited diversity, which invalidates meaningful market segmentation. We also point out that based on music fans, the only determining factor that separates music artists is their popularity rather than their music styles etc. , which from another perspective invalidates the "Fans Economy" assumption.

We collect our data from the B company (For commercial privacy issues, we would like to keep the anonymity). B company is one of the largest Chinese online music sites with millions of unique visitors per month. On the website, the user could listen to music, download music , search for music or join a discussion board of a specific music artist to interact with other fans online.

The artist discussion board forms a large artist-fan social network worthy of research study . In the following sections we provide social network analysis of this artist-fan social network in the hope to provide insight to the Chinese music market. In the meantime, we point out the "Fans Economy" assumption of the Chinese popular music industry is invalid.

## II. RELATED WORK

Social Network Analysis has been attracting researchers' and industrial engineers' attention since the emergence of SNS websites such as Facebook and Twitter. SNS websites have accumulated large amounts of users' social interaction information, which made social network analysis based on big data possible. Facebook even created a data department with devoted effort on social network analysis. They have done a series of insightful research on Facebook data. For instance, they discovered Facebook network has a four degrees of separation [1] . They have also used social network information for other applications such as predicting user's geographical location [2]. Other companies like Renren.com [3] have also done extensive study on social networks.

| 张艺兴(EXO) 34018 | 冯建宇 26841 | EXO 9386 | TFBOYS 7421 | 防弹少年团 7873 | Taylor Swift 4281 |
|---|---|---|---|---|---|
| EXO 7038 | 王青 16597 | Super Junior 871 | TF 家族 914 | BigBang 5735 | 陈奕迅 3273 |
| 鹿晗 7016 | 许魏洲 310 | 圭贤(Super Junior) 701 | 庄心妍 540 | G-Dragon(BigBang) 3750 | Avril Lavigne 2330 |
| 边伯贤(EXO) 6120 | 黄景瑜 288 | 东海(Super Junior) 684 | 智妍(T-ara) 539 | f(x) 3636 | 张学友 2235 |
| 吴世勋(EXO) 4585 | 陈秋实 56 | 银赫(Super Junior) 680 | T-ara 530 | 少女时代 3300 | 王菲 1782 |
| 朴灿烈(EXO) 4580 | 赵泳鑫 40 | 强仁(Super Junior) 540 | 王俊凯 518 | EXO 2550 | 刘德华 1781 |
| EXO-K 4152 | 蔡照 40 | Super Junior-K.R.Y 539 | 易烊千玺 424 | T.O.P(Bigbang) 2196 | 王力宏 1663 |
| EXO-M 3994 | 檀健次 38 | 利特(Super Junior) 535 | 王源 408 | 太妍(少女时代) 1912 | 少女时代 1614 |
| Chen(EXO) 3459 | MIC 男团 37 | 神童(Super Junior) 528 | 鹿晗 92 | 太阳(Bigbang) 1865 | 周杰伦 1377 |
| 都暻秀(EXO) 3432 | 池约翰 33 | 始源(Super Junior) 525 | EXO 81 | 胜利(Bigbang) 1728 | S.H.E 1303 |

Table 1 Most popular artists of the 6 largest communities in Fan-Fan Network

In this paper, to justify "Fans Economy" does not work for the Chinese pop music industry, we utilize the community detection technique. Community detection is a well studied topic with extensive research literature [4] [5][6]. In this paper, we resort to method proposed in [4] for market segmentation. We prefer the method because it has been well tested for years and it is well integrated with Gephi, which provides easy-to-use and manipulatable visualization of social networks.

"Fans Economy" has been popular in China in recent years . L. Tie and F. Yong [7] points out consumers have a sense of being the producer when they become loyal fans of a product and interact with the real producer of the product. K.Ye [8] claims "Fans Economy" is crucial for a conventional company to successfully transform itself into an internet company. IT companies like Xiaomi, Lenovo and music entertainment website like iQiYi are all followers of the concept of Fans Economy.

III. SOCIAL NETWORK ANALYSIS

A. Degree Distribution

We introduce the following Artist-Fan network G(V, E) . In the network, there are two types of vertices - the first type is artist vertices, each of which represents an artist; the second is fan vertices, each of which represents a fan. If a fan joins the discussion board of an artist , then a directed edge from the fan to the artist is formed. There are no edges between artists. The Artist-Fan network from out data set contains data from the year 2016 with 13054 artist and 660054 fans.

Figure 1 shows the log-log plot of the out-degree distribution of fan vertices and Figure 2 shows the log-log plot of the in-degree distribution of the artist vertices. Both distributions are approximately power-law distributions.

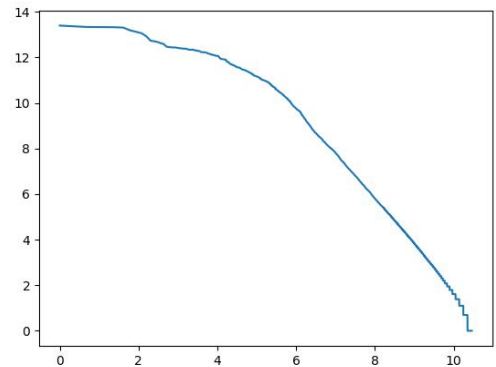

Figure 1. Log-log plot of the number of fans following a music artist

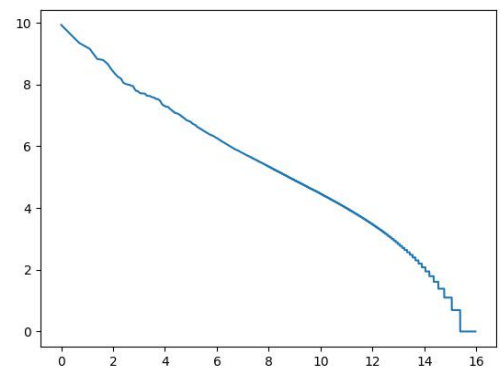

Figure 2. Log-log plot of the number of artists that a fan follows

The power law distribution effect indicates that only the few most popular music artists gain massive popularity. The rest of them are much lesser known to the public. Similarly, the majority of popular music consumers follow a small-to-moderate number of artists.

## B. Community Detection

We create the following Fan-Fan social network G(V, E) where each vertex represents a music fan. A weighted edge is formed between two vertices if two music fans join the discussion board of the same artist in the year of 2016. The edge weight represents how many music artists they share. To simplify our computation, we omitted edges whose weights are smaller than 3. In the end, we obtain an undirected graph of 152598 vertices and 38282018 edges.

We computed the modularities of Fan-Fan network using the multi-level Louvain method [4] shipped with the "igraph" package of R. The community detection algorithm generates 83 communities with 6 dominating communities and 77 smaller communities of negligible sizes [Fig 3].

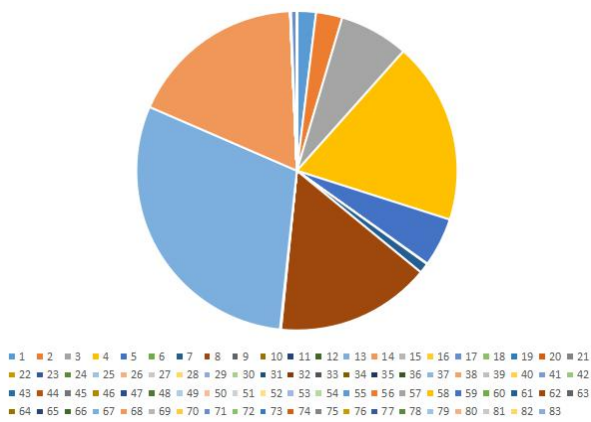

Figure 3. Community Sizes of Fan-Fan Network produced by Louvain Method

We compute the most popular artists in each of the 6 largest communities and show the result in Table 1. From the statistics, in 2016 the most popular music artist is Korean bands EXO, with a couple of Chinese singers coming next. It is obvious the popular artists from Hong Kong and Taiwan are no longer popular as they used to be.

The Matthew effect of each community is very obvious. Most fans in each community follows one or two of the most popular artists with the exception of the smallest community. For example, in the first community of Table 1, the top 10 most popular artists are all from the Korean band EXO. The number of fans following other artists are much fewer.

This observation could be very frustrating to "Fans Economy" believers because in the music industry, the Matthew effect is so strong even in segmented communities. You could not create many artists with self-exclusive fan bases. **The game rule is you either introduce an artist that is the most popular or you get almost no fans at all.**

In addition, there aren't so many different music styles a music marketer could pick for a newly introduced music artist. Out of the 6 largest communities in Fan-Fan Network, 3 are Korean, 1 is Hong Kong and Taiwan that has no great prospects. For the other 2 communities, the young Chinese boy band TFBoys dominates 1 community, the Chinese artists Jianyu Feng and Qing Wang dominate the other community. For a Chinese music marketer, you could either import a popular Korean band into the Chinese market, or you introduce a local artist similar to TFBoys or Jianyu Feng / Qing Wang. **The Chinese pop music market does not have enough sub-markets for "Fans Economy" to work.**

Now let's take a look at the popular music artists: We consider the following Artist-Artist social network G(V, E) where each vertex represents a music artist. A weighted edge is formed between two vertices if two music artists share a music fan on both of the artist discussion board in the year of 2016. The edge weight represents how many music fans they share. We delete edges whose weights are smaller than 2 and get a social network of 923 vertices and 425503 edges.

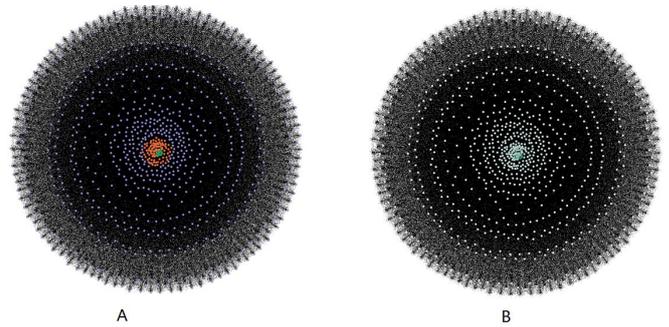

Figure 4 Community Detection of the Artist-Artist network with resolution 1.0 (A) Community Detection (B) PageRank Values

We compute the modularities of Artist-Artist network using algorithm introduced in [4] with Gephi. On appearance, the social network has well formed modularity structures. When resolution is 1.0, we obtain 3 communities. The communities comprise of 87.65%, 9.97% and 2.38% of the vertices. We also compute the weighted PageRank values of the vertices. After comparison, we find out the community detection result coincides with segmentation based on weighted PageRank values. In other words, vertices of the largest 2.38% PageRank values form the first community, 9.97% of the next largest PageRank values form the second community, the rest of the vertices form the third community. We also try the community detection algorithm when resolution is smaller. However, even when resolution is smaller, we get similar results to when resolution is 1.0, i.e., one dominating community and coincidence with PageRank Values.

The community detection result of the Artist-Artist network demonstrates that **based on music fans, the pop music**

**artists could only be segmented by popularity.** This in turn debunks the myth that "Fans Economy" works for the Chinese popular music industry because based on fans, the only determining factor that segments artists is not their music styles or genres but their popularity.

## IV CONCLUSION

In this paper, we apply social network analysis to debunk the popular myth in the Chinese popular music industry that "Fans Economy" is an ideal model for new artist introduction. We demonstrate that popularity and Matthew effect invalidates the attempt to create exclusive fans for a specific music artist based on factors other than popularity. We also demonstrate that the diversity of the sub-markets of Chinese popular music industry is highly limited and unsuitable for market segmentation.